\documentclass[<2p>]{article}
\usepackage{graphicx}
\usepackage{amssymb,amsfonts,amsmath}
\usepackage{color}
\usepackage{authblk}
\usepackage{etoolbox}
\patchcmd{\thebibliography}{\section*{\refname}}{}{}{}
   
\begin{document}
\date{}
\title{A new failure mechanism in thin film by collaborative fracture and delamination: interacting duos of cracks}

\author[1,2]{Jo\"{e}l~Marthelot\thanks{joel.marthelot@espci.fr}}
\author[1]{Jos\'{e}~Bico}
\author[3]{Francisco~Melo}
\author[2]{Beno\^{i}t~Roman}
\affil[1]{\small PMMH, ESPCI-PSL, CNRS, UPMC Paris 6, UPD Paris 7, Paris, France}
\affil[2]{SVI, CNRS UMR 125, Saint-Gobain Recherche, Aubervilliers, France}
\affil[3]{Departamento de F\'isica, Universidad de Santiago de Chile, Santiago, Chile}

 
\maketitle

\begin{abstract}
When a thin film moderately adherent to a substrate is subjected to residual stress, the cooperation between fracture and delamination leads to unusual fracture patterns, such as spirals, alleys of crescents and various types of strips, all characterized by a robust characteristic length scale. 
We focus on the propagation of a duo of cracks: two fractures in the film connected by a delamination front and progressively detaching a strip.
We show experimentally that the system selects an equilibrium width on the order of 25 times the thickness of the coating and independent of both fracture and adhesion energies. We investigate numerically the selection of the width and the condition for propagation by considering Griffith's criterion and 
the principle of local symmetry. In addition, we propose a simplified model based on the criterion of maximum of energy release rate, which provides insights of the physical mechanisms leading to these regular patterns, and predicts the effect of material properties on the selected width of the detaching strip.
\end{abstract}

\section{Introduction}

\begin{figure}
\centering
  \includegraphics[width=11cm]{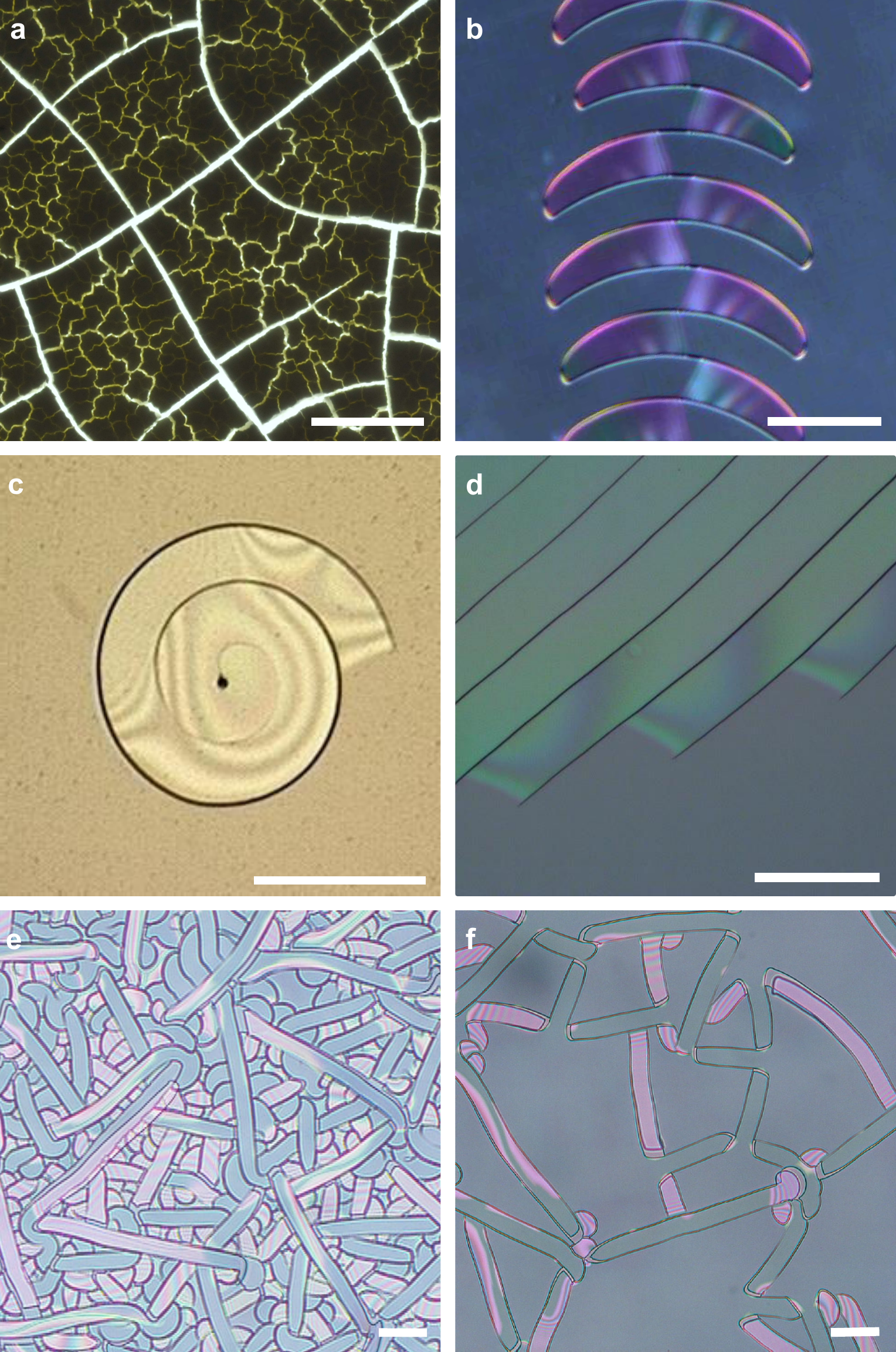}
  \caption{Cracks in thin films. The standard propagation of isolated cracks in a thin drying layer of corn starch leads to a hierarchical network where new cracks branch perpendicularl to older ones (a, $h_f=2.7~$mm, scalebar 1~cm). However, surprising regular patterns are observed in thin silicate (``Spin-On-Glass'') layer when the adhesion of the coating is moderate (scalebar $100~\mu$m): long oscillating alleys (b, $h_f=1.2~\mu$m, $\Gamma_i^0=0.45$~N/m), Archimedean spirals (c, $h_f=1.1~\mu$m, $\Gamma_i^0=0.3$~N/m) or parallel cracks (d, $h = 1.45~\mu$m, $\Gamma_i^0=0.3$~N/m) follow an initial arch or line, at a fixed distance. Denser defects lead to more complex patterns, nevertheless the band structure remains robust (e, f).}
  \label{Fig:MTEOS}
\end{figure}

Thin film coatings are commonly used in engineering applications as a way to enhance the surface properties of a substrate.
However, residual stresses generally build up in the layer during deposition or use, which may result into  buckle-driven delamination when stresses are compressive or fracture in the opposite case of tensile stresses (see the classical review from Hutchinson and Suo~\cite{Hutchinson92}).
In this paper we focus on this later situation, which has been widely explored due to the dramatic consequences of such a failure. 
In the case of a strong adhesion of the thin film, the condition for the propagation of {\it channel cracks} is well documented and we summarize the main results in section~\ref{channel}.
When the propagation of cracks is possible ({\it i.e.} beyond a critical tensile load), cracks usually appear sequentially: a new crack follows a straight trajectory, until it deflects and connects perpendicularly to an earlier crack as a consequence of the tensorial nature of the strain (fig.\ref{Fig:MTEOS}a). 
If the substrate is rigid, the length scale of interaction between cracks  is on the order of the thickness of the layer. Indeed, the substrate tends to limit the relaxation of the residual strain induced by fracture~\cite{Thouless90, Thouless92, Shenoy00}. The resulting hierarchical fracture patterns appears in old paintings~\cite{Karpowicz1990}, drying mud~\cite{Shorlin00} or sol-gel coatings~\cite{Atkinson91}. In addition to channel cracks, delamination is also observed for high tensile stresses. The interplay between channel cracks, delamination and substrate fracture has motivated a large number of studies~\cite{Cannon85, Mei07, Yu03}, to provide the operational conditions for thin film stability~\cite{Ye92, Evans95, Freund04}. However, delamination requires free edges to propagate. As a consequence, this mode of failure is usually  observed between previously formed channel cracks and eventually leads to the detachment of pieces of the coating~\cite{pauchard06, Lazarus11}. 

The nucleation of successive channel cracks is nevertheless not the sole scenario for the failure in coatings under tensile stresses.
Indeed, a different mode has been recently evidenced in systems  involved in new applications such as stretchable electronics~\cite{Rogers10}, biomaterials~\cite{Haldar06} or fuel cells~\cite{Bozzini12}, which display a weak adhesion of a thin film on a substrate.
Intriguing fracture patterns have been reported in such systems such as regular alleys of oscillating crescents (fig.~\ref{Fig:MTEOS}b), spiraling paths (fig.~\ref{Fig:MTEOS}c), parallel cracks (fig.~\ref{Fig:MTEOS}d), or more disorganized patterns (fig.~\ref{Fig:MTEOS}e)~\cite{Sendova03, Meyer04, Lebental07, Wan09, Wu13}. 

As a mean feature, these patterns display a well defined characteristic length scale, which is best evidenced in the wavelength of the crescent alleys (fig.~\ref{Fig:MTEOS}b) or the pitch of the Archimedean spirals (fig.~\ref{Fig:MTEOS}c). 
This lengthscale is typically 25 times the thickness of the coating for most experimental systems.
In a recent paper presenting live observations of crack propagation, we showed that these striking patterns result from the simultaneous propagation of a delamination front in collaboration with fracture across the layer~\cite{Marthelot14}. 
This new collaborative fracture mode surprisingly develops in conditions where standard channel cracks are not expected to propagate.
We provided a simplified physical model based on energy estimates that selects a length scale for the patterns proportional to the thickness of the film. 
This model also reproduces the diversity of observed patterns. 
However, a quantitative prediction of the crack propagation as a function of the material properties of the coating and the substrate is still missing. 
We focus here on the case of a symmetric pair of cracks propagating simultaneously with a delamination front and leading to a strip of a well defined width (fig.~\ref{Fig:MTEOS}f and~\ref{Fig:W_vs_h}a).
Indeed, this simpler geometry allows for a quantitative modeling  and a direct comparison with experiments.
Studying this configuration thus constitutes a first step towards a generalization to more complex modes such as spirals or crescent alleys.

In section~\ref{section:duo}, we characterize experimentally symmetric pairs of cracks and show how the selected width depends on the different physical parameters.
We address in section~\ref{section:K2} the selection of this width by studying numerically the stress intensity factors at the crack tips, from which the direction of propagation of the cracks can be predicted. We also determine the conditions of propagation using a steady-state argument.
In section~\ref{section:NRJ}, we propose a complementary approach based on maximum energy release rate \cite{Chambolle09}, which provides insights into the physical mechanisms that select the width of the band.
In section~\ref{section:sublayer}, we examine the robustness of the selection of the width and determine its dependence with material properties (Poisson ratio, Young modulus of the layer and of the substrate) using three-dimensional finite element computations and simple analytical arguments based on energy. We finally quantitatively compare those predictions with experimental data. 

\section{Experimental evidence of a new mode of crack propagation}
\label{section:duo}

``Spin-On-Glass" silicate coatings (SOG) are commonly used as protective coatings or to adjust the dielectric properties of materials~\cite{Lebental07}. 
Depositing such films relies on a simple process where a sol-gel suspension is spin-coated on a substrate and then cured in an oven. 
Nevertheless, the sol-gel condensation reaction produces residual tensile strains (on the order of 1\%), which may induce fracture~\cite{Brinker90}. 
When adhesion with the substrate is large, such stresses indeed induce the propagation of isolated channel cracks, leading to the usual hierarchical pattern of glaze cracks in ceramics (fig.~\ref{Fig:MTEOS}a).
However, if the adhesion of the layer on the substrate is moderate, unusual crack patterns can also appear as illustrated in fig.~\ref{Fig:MTEOS}.
Although similar patterns have been reported with other materials, we selected SOG coatings as a model system for controlled experiments. 

\subsection{Experimental setup}
\label{exp_setup}
\begin{figure}
\centering
  \includegraphics[width=12cm]{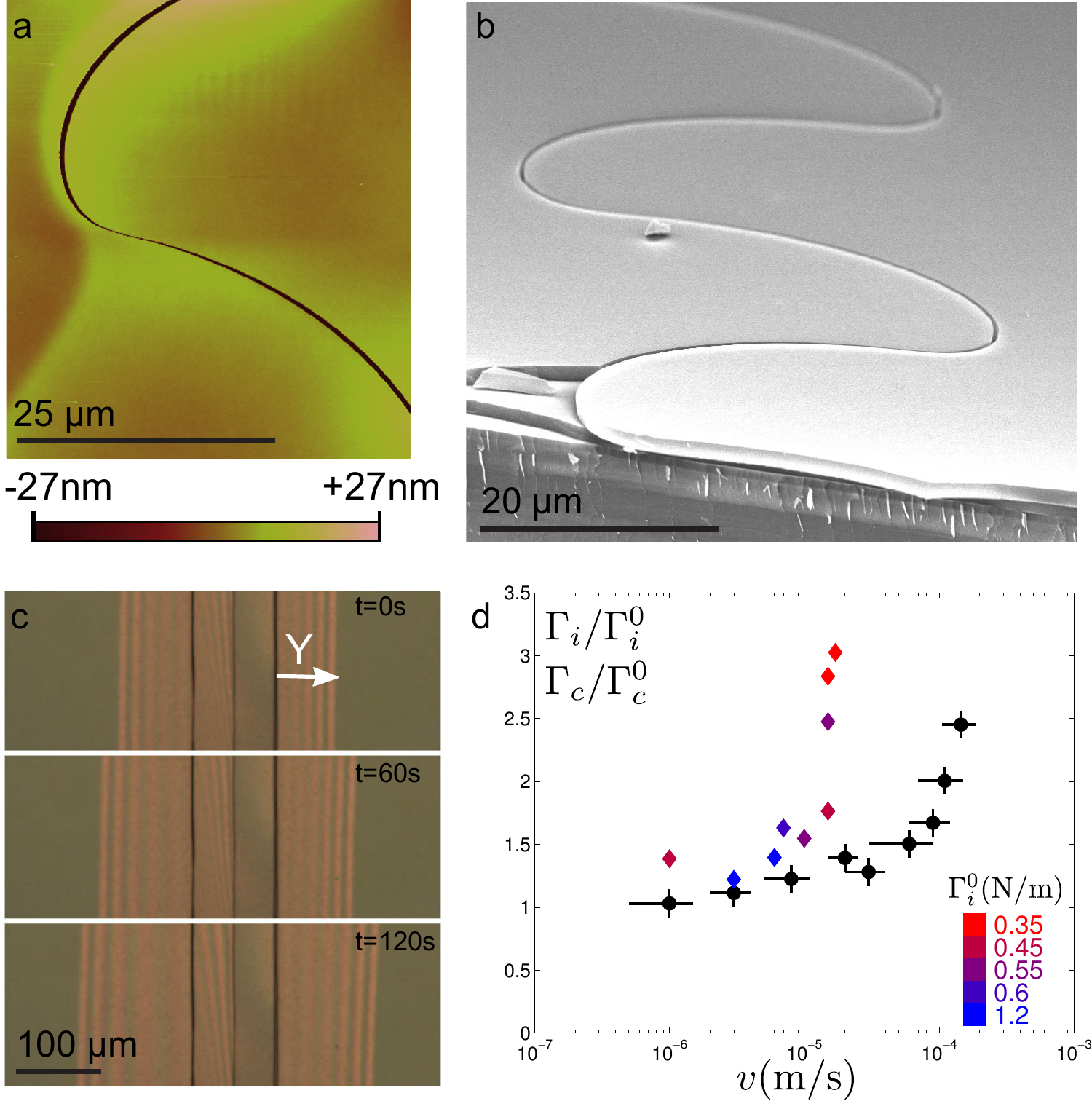} 
     \caption{(a) AFM imaging of crescents alleys reveals a slight increase of the thickness of the silicate layer as residual strain is relaxed by the crack. 
(b) Field Emission Gun-SEM images of the crack path. (c) Evolution of a delamination front around a fixed straight channel crack. (d) The delamination energy $\Gamma_i$ estimated from the shape of the front (diamonds) increases with the velocity. The delamination energy $\Gamma_i$ is normalized by the value of the delamination at zero velocity $\Gamma_i^0$ (coded in colors). Similarly, the fracture energy $\Gamma_c$ measured in isolated cracks (black dots) increases with the crack velocity.
}
  \label{fig:properties}
\end{figure}

Thin films were produced by spin-coating a commercial solution of organosilicate sol-gel (SOG spin-on-glass, Accuglass T-12B, Honeywell) on a silicon wafer. 
The rotation speed was set from 500 to 1200 rpm for 15 to 25 seconds. 
The liquid layer was placed in an oven at 200$^\circ$C for 2 hours. 
The solvent then quickly evaporated and the sol-gel condensation reaction progressively transformed a gel into a glassy micrometric silicate film. 
The obtained silicate coating was characterized by different parameters, that we have quantified precisely, namely the thickness $h_f$, Young modulus $E_f$ and the Poisson coefficient $\nu_f$ of the film, the residual strain $\sigma_0$, the adhesion energy $\Gamma_i$ and the fracture energy $\Gamma_c$.\\
The thickness $h_f$ was measured by AFM (fig.~\ref{fig:properties}a) or SEM imaging of gently scratched specimens (fig.~\ref{fig:properties}b). 
The Young modulus $E_f=4\pm 1$ GPa was estimated by standard nanoindentation (MTS XP from Agilent).
The Poisson coefficient was inferred from the measurement of the relative variation of the film thickness $\epsilon_{zz} = 2\frac{\nu_f}{1-\nu_f}\epsilon_{xx}$ induced by the delamination of a strip of the film.
This variation on the order of 6~nm for a film of $h_f = 1\,\mu$m was measured by AFM imaging, corresponding to $\nu_f=0.25$ (fig.~\ref{fig:properties}a).\\
The tensile stress was obtained by monitoring the slight deflection from the slight deflection induced on the supporting wafer (we used a wafer of thickness $h_s = 100\,\mu$m for these measurements). This measurement confirmed that residual stresses are isotropic in the film.
The curvature of the wafer $\kappa$ is indeed related to the residual stress through Stoney's law, 
$\sigma_0=\frac{E_s h_s^2 \kappa}{6h_f(1-\nu_s)}$, 
where $E_s=169\,$GPa, $\nu_s = {0.36}$ are respectively the Young modulus and the Poisson coefficient of the silicon wafer~\cite{Stoney1909}. We typically obtained $\kappa\sim 0.15\,\mathrm{m}^{-1}$ for $h_f=1\,\mu$m, leading to a residual stress
$\sigma_0 = 55\,$MPa for the different samples.
Such stress corresponds to a residual strain $\epsilon_0 = \sigma_0/E_f$ on the order of $1\%$. \\
As a way to control the adhesion of the coating layer, the wafer could be previously coated with a first thin layer of SOG, methyltriethoxyorthosilicate, tetraethylorthosilicate, glycidoxypropyltrimethoxysilanetetraethoxysilane or 1H,1H,2H,2H perfluorodecyltrichlorosilane.
The adhesion energy $\Gamma_i$ was inferred by monitoring the shape of a delamination front around a fixed straight crack of length $L$ as proposed by Jensen {\it et al.}~\cite{Jensen90} (fig.~\ref{fig:properties}c). 
The aspect ratio of the front $Y/L$ is indeed predicted to rely on the ratio of the residual stress $\sigma_0$ with a critical stress $\sigma_c$, on the Poisson ratio of the film and on a parameter accounting for mode 3 contribution at the tips of the straight crack. 
However, the effect of this last parameter is quite moderate and the numerical data from Jensen {\it et al.} 
computed for $\nu_f = 1/3$ and different contributions of mode 3 can be fitted by the simple relation $\sigma_0/\sigma_c = 1+ 0.47(\pm0.05)Y/L$.
Measuring the critical strain for delamination from the shape of the front thus provides an estimate of the adhesion energy through the relation $\sigma_c = \left(\frac{2E_f\Gamma_i}{(1-\nu_f^2)h_f}\right)^{1/2}$~\cite{Jensen90}.
As it is observed in any adhesive system, the adhesion energy increases with the speed of propagation of the delamination front (fig.~\ref{fig:properties}d). This is not due to kinetic energy effects (typical velocities range from 1 to 100 $\mu$m/s) but to time dependent dissipation ({\it e.g.} diffusion of reactive molecules facilitating 
fracture propagation~\cite{Lin07}). We refer to as $\Gamma_i^0$ the limit of this energy estimated for a motionless front. 
By changing the preliminary coating on the silicon wafer, we were able to vary $\Gamma_i^0$ between 0.3 and 1.3 N/m.\\
A standard plasma cleaning of the sample finally promotes a strong adhesion and thus prevents delamination.  
In this condition, channel cracks are observed above a critical thickness $h_c$, as expected in thin films under tension (see section~\ref{channel}).
For a residual stress $\sigma_0 = 55\,$MPa, a critical thickness $h_{c}= 1.8\,\mu$m 
 lead to a fracture energy, $\Gamma_c^0=1.5 \pm 0.2\,$N/m according to equation~(\ref{eq:hc}).
Fracture energy was also found to increase with the speed of propagation of the crack (fig.~\ref{fig:properties}d). This dependence is compatible with observations in other similar systems where it has been interpreted as effect of water diffusion through the fracture tip~\cite{Lin07}.

\subsection{Duos of cracks: experimental observations}

\begin{figure}
\centering
  \includegraphics[width=12cm]{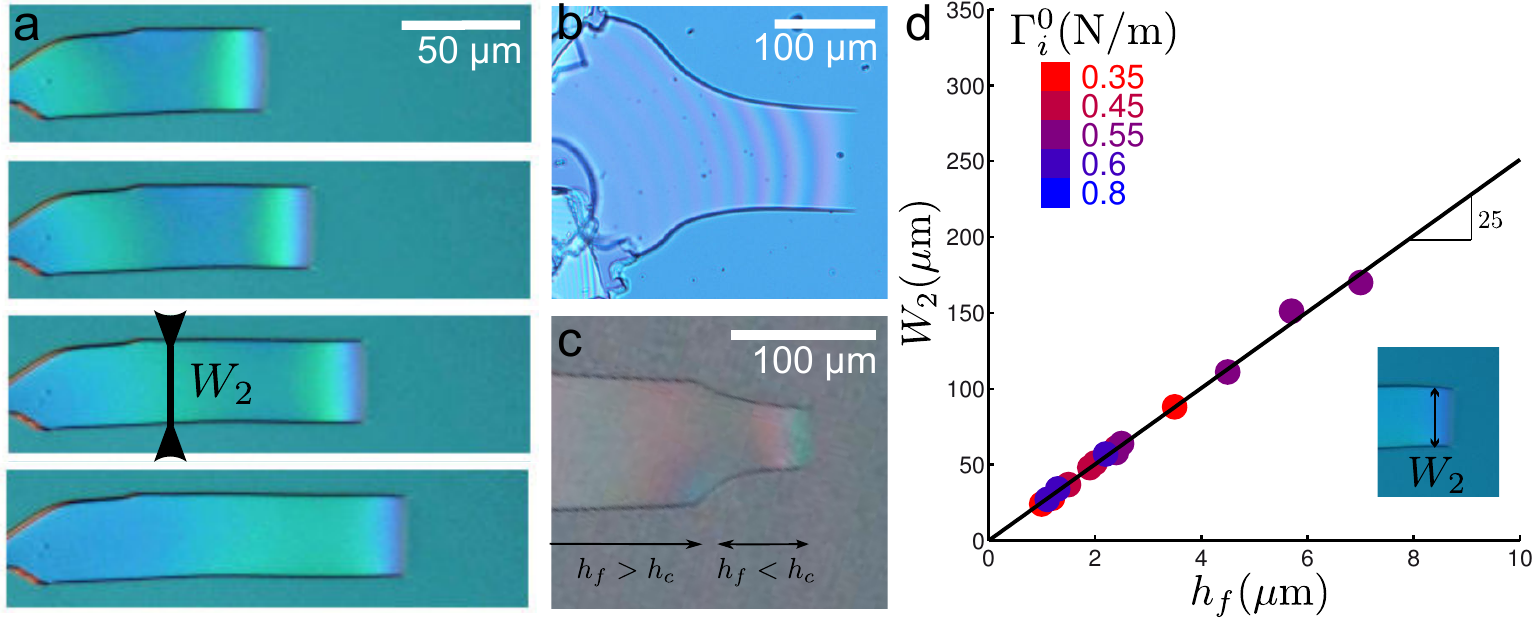}
  \caption{(a)  Propagation of duos of cracks in a thin film of spin-on-glass, deposited on a silicon wafer ($h_f=1.4\mu m$, $\Gamma_i^0=0.55$ N/m).  
  In this particular example, the cracks initially linked by a narrow delamination front, start to diverge but quickly converge towards a steady width $W_2$. The pictures are taken with a constant time lap of 2 s.
  (b) Opposite case where the initial width is larger than $W_2$, leading to the same optimal width.
 (c) Relaxation of a duo of cracks towards the optimal width in a film presenting a gradient in thickness. 
   In the left part of the image, the thickness is larger than the critical thickness $h_c$ required for the propagation of an isolated channel crack. Two isolated cracks propagated independently and stopped as they reached the region below $h_c$. 
   A slow delamination front then propagates between both cracks and the collaborative propagation starts, leading to a rapid relaxation towards $W_2$. 
(d) Variation of $W_2$ as a function of the film thickness $h_f$ for different values of the adhesion energy $\Gamma_i^0$.}
\label{Fig:W_vs_h}
\end{figure}

Although a large variety of patterns can be observed for different triggering conditions~\cite{Marthelot14}, we focus in this paper on the paradigm case of duos of cracks that propagate simultaneously. 
Such pairs were for instance observed to grow perpendicularly along scratches produced with a sharp blade.
Optical interferences visible as the pair of cracks advances, indicate that the film delaminates simultaneously (fig.~\ref{Fig:W_vs_h}a). 
Although the initial width of the delaminated band is set by the initial conditions, it 
systematically converges toward a steady value $W_2$, which is found constant for a sample of uniform thickness. 
In this steady regime, the delamination front connecting both cracks propagates at a constant speed ranging from 1 to $100\,\mu$m/s.

We measured the variation of the stationary width of the strips $W_2$ as a function of $h_f$ for different adhesion energies $\Gamma_i^0$. As illustrated in fig.~\ref{Fig:W_vs_h}d, $W_2$ increases linearly with the thickness over one decade. We observe that the duos of cracks converge rapidly towards $W_2$ (fig.~\ref{Fig:W_vs_h}b and c), which is in the order of 25 times the film thickness. 
This prefactor which probes the interaction distance between the cracks is strikingly large in comparison with isolated cracks where the interaction length is on the order of $h_f$. Moreover, this optimal width is surprisingly robust and does not depend on the actual value of the adhesion energy $\Gamma_i^0$.
In the remaining of this article, we wish to explain these two remarkable experimental facts, through two different approaches.

\section{Prediction of crack propagation through stress intensity factors}
\label{section:K2}

In a first approach, we study the stress field around the crack tips, using the standard tools of linear elastic fracture mechanics and finite elements numerical calculations. In this section (as well as in section~\ref{section:NRJ}), we restrict the analysis to the case of a single layer on a infinitely rigid substrate. A quantitative comparison with experiments including a sublayer is presented in section~\ref{section:sublayer}.
Before describing duos of cracks, we start the following section by reformulating standard results on isolated cracks~\cite{Hutchinson92}. 

\subsection{Standard isolated cracks}
\label{channel}

Consider the classical case of a single and straight quasi-static fracture propagating in a thin isotropic homogeneous coating perfectly adhering to an infinitely rigid substrate. We note that in view of the deposition technique (condensation of a layer) the films in our experiments have in-plane isotropic mechanical properties, and undergo isotropic residual stresses, as confirmed by direct measurements described in section~\ref{exp_setup}. The questions that we wish to address are as follows: what are (i) the direction of propagation and (ii) the conditions for propagation?  

The calculation of the stress field close to a crack tip in the film is a complex three-dimensional problem. 
However, the configuration is mirror symmetric with respect to the straight crack trajectory, so that the stress field must also be symmetric (the loading is in pure opening mode I) all along the fracture front.
According to the principle of local symmetry, the direction of propagation is therefore in the plane of the initial trajectory: (i) an initially straight crack path remains straight. 
Note that this symmetry argument does not state on the stability of such an existing straight trajectory: a perturbation to this solution may be amplified 
during propagation. 
The symmetry of the system can be broken through a bifurcation. Indeed oscillating crack path is for instance observed in plates subjected to a thermal gradient~\cite{Yuse93, Deegan03} or during the cutting of a thin sheet with a blunt tool~\cite{ghatak03, Roman03, Audoly05}, two situation where the
system is mirror symmetric with respect to a straight path. In the case of interest here, we assume straight paths to be stable, in the standard sense of stability analysis, since they are experimentally observed.

\begin{figure}
\centering
  \includegraphics[width=12cm]{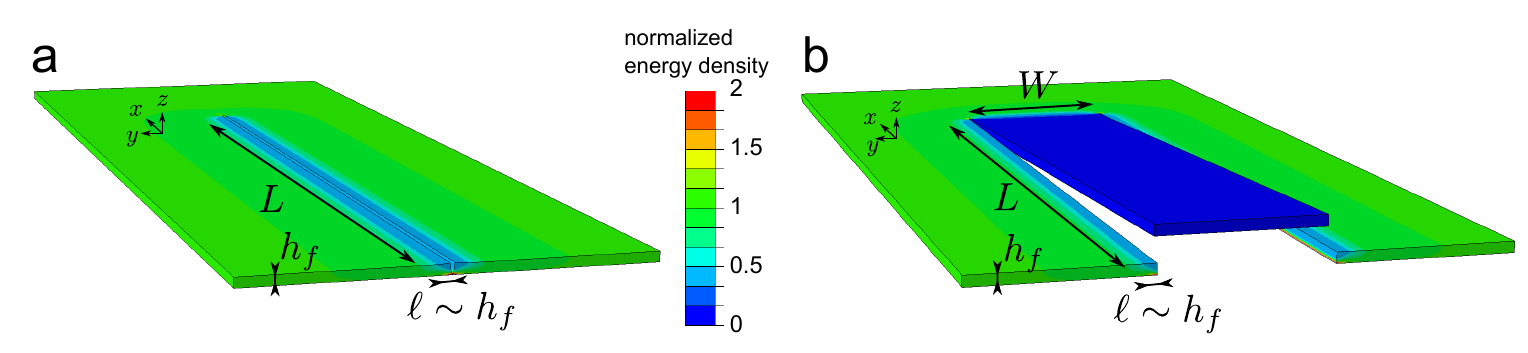}
  \caption{(a) Elastic energy distribution around an isolated channel crack in a thin layer under residual stress $\sigma_0$. The energy density is normalized by its value in absence of fracture $e/h_f=\sigma_0^2(1-\nu_f)/E_f$. Elastic energy is released in a small area around the crack edges, with typical lateral size $\ell \sim h_f$ in the direction $y$ perpendicular to the crack. (b) Collaborative mode of propagation of a duo of cracks separated by a delamination front. In addition to the lateral strain release along crack path, the strip defined by three free edges is almost completely free of elastic energy in this propagation mode.}
  \label{Fig:Collaborative}
\end{figure}

In addition to the shape of the path, we can determine the condition for propagation through Griffith's criterion. 
The energy release rate of an isolated crack (fig.~\ref{Fig:Collaborative}a) is obtained from a simple plane strain analysis instead of the resolution of the three-dimensional problem in the vicinity of the crack tip. Indeed, the difference in energy between a section far ahead and far behind the crack tip leads to the following energy release rate $G$: 
\begin{equation}
G h_f = 2 \gamma e h_f
\label{eq:Beuth}
\end{equation}
where $e=h_f\sigma_0^2(1-\nu_f)/E_f$ is the elastic energy density in the film per unit area, and $2\gamma$ is a numerical prefactor which depends on the Poisson ratio $\nu_f$ of the film.
We have noted $h_f$ the film thickness, $E_f$ its Young modulus and $\sigma_0$ the residual stresses in the film.
Numerical tabulations for $\gamma$ can be extracted from~\cite{Beuth92}, where the effect of a finite
rigidity of the substrate is also considered through the Dundur parameters~\cite{Hutchinson92}. In the case of an elastic coating under residual stress deposited on a rigid substrate, $\gamma$ is of order 1 (we used $\gamma=0.64$ in our measurement of fracture energy in section \ref{section:duo}), but can be much larger for compliant substrates. We will consider such effects in section~\ref{section:quantitative}.

Following Griffith criterion, isolated channel cracks propagate quasi-statically (ii) in thin films if the energy release rate reaches the fracture energy $\Gamma_c$. 
\begin{equation}
 2 \gamma e  = \Gamma_c  
 \label{eq:Griffith_isolated}
\end{equation}
This condition is usually presented in the following form: for a given residual strain $\sigma_0$, the film is stable with respect to fracture ({\it i.e.} a nucleated crack will not propagate) if the film thickness $h_f$ is smaller than a critical value $h_c$:
\begin{equation}
h_c=\frac{\Gamma_c E_f}{2 \gamma \sigma_0^2 (1-\nu_f)} 
\label{eq:hc}
\end{equation}
In our experiments such isolated cracks are observed  in sufficiently thick coatings when large adhesion prevents delamination. As described in section~\ref{exp_setup}, we used measurement of the critical thickness $h_c$ to estimate the work of fracture $\Gamma_c$.

These results are well established~\cite{Hutchinson92,Thouless90,Thouless92,Shenoy00,Beuth92}, and recent studies focus on the interaction between several isolated cracks. A simplified model based on linear foundation connecting the layer to the substrate has for instance been proposed~\cite{Xia00} and  later rigorously derived and used to study the sequence from fracture to debonding~\cite{Baldelli13, baldelli14}.

\subsection{Predicting the width separating duos of cracks}
\label{duosKII}

We now focus on the case of duos of cracks, which is not described by the classical situation presented before.
We nevertheless follow the same line of arguments for this new collaborative mode: we first study the stress field around the cracks to determine the path and then use Griffith criterion to predict the condition for propagation.

Consider a duo of cracks propagating with a steady width $W_2$ as illustrated in fig.~\ref{Fig:W_vs_h}a.
The thin film initially adheres to a rigid substrate and two straight cracks propagate through the film, with the resulting strip simultaneously delaminating from the substrate. 
In contrast with the case of isolated cracks, the mechanical conditions on each side of the cracks tips are very different.
Indeed, the film remains attached to the substrate in one side and free in the other side.
As consequence there is no {\it a priori} reason to expect a pure mode-I loading and therefore a straight propagation of the cracks. However, fracture may experience pure opening loading even in configurations where there is no symmetry, for instance in the spalling and cracking processes of brittle plates~\cite{Hutchinson92,Thouless87} or in the case of the interaction of curved channel cracks~\cite{Xia00}.

Finite-element simulations of a straight delaminated strip were performed within the limit of three-dimensional linear elasticity with the Abaqus software from Dassault Syst\`emes.
The mesh size used for these simulations was one tenth of the film thickness. For the sake of simplicity, we first consider the case of a layer
directly deposited on an infinitely rigid substrate (fig.~\ref{Fig:Collaborative}b).
The effect of the finite Young modulus of the substrate or that of a sublayer will be later described in section~\ref{section:sublayer} for a quantitative comparison with experimental data.
The system is therefore composed of an isotropic thin film  (thickness $h_f$, Young modulus $E_f$, Poisson ratio $\nu_f$) in which two rectilinear cuts are performed, 
delimiting a rectangular strip with a width $W$ (fig.~\ref{Fig:Collaborative}b). 
The layer is attached to the infinitely rigid substrate except in the area below the strip. We use the symmetry of the problem to simulate only half of the strip.
The isotropic biaxial residual stresses are imposed in the numerics as the result of a thermal contraction of the coating.

We use two simplifying assumptions on the geometry of the fracture front, which are guided by experimental observations. 
A first assumption is that the debonding front is straight as observed experimentally (see images on fig.~\ref{Fig:W_vs_h}a).
A second simplification concerns the shape of the channel crack in the thickness of the coating.
Post-mortem SEM imaging suggests that the fracture plane is nearly perpendicular to the film surface (fig.~\ref{fig:properties}b).
Although a precise monitoring of the crack front is lacking, we will assume that the fracture front in the film is straight and perpendicular to the film surface.

\begin{figure}[h!]
\centering
  \includegraphics[width=12cm]{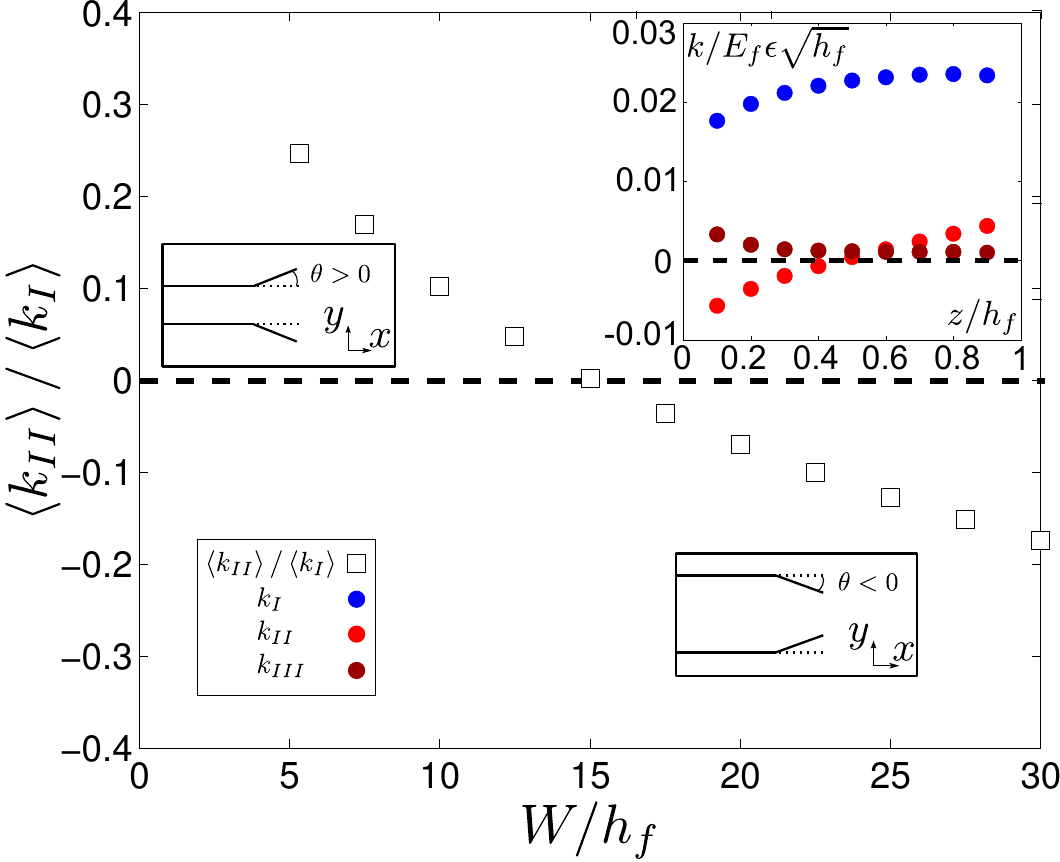}
  \caption{Numerically computed stress intensity factors at the crack tips for a delaminated strip with width $W$ in a film subjected to residual bi-axial stress. 
The ratio $\left< k_{II} \right>/ \left<k_{I}\right>$ is plotted as a function of the normalized width $W/h_f$, where $h_f$ is the thickness of the coating (stress intensity factors are averaged in the thickness of the film). 
 Since the sign of the ratio is related to the direction of propagation of the crack, parallel cracks are expected for $W \sim 15.1 h_f$. Inset : Profiles of $k_I$, $k_{II}$ and $k_{III}$ along the crack front $z/h_{f}$ for the selected width $W_2 \sim 15.1 h_f$. The position $z/h_{f}=0$ corresponds to the lower surface of the film in contact with the substrate and $z/h_{f}=1$ to the upper surface.
}
  \label{Fig:Numeric_K2}
\end{figure}

Stress intensity factors were computed for both cracks  across the thickness of the film for different widths $W$ of the strip. 
The averaged values of the stress intensity factors over $h_f$ are respectively quoted as $\left<k_I\right>,\left<k_{II}\right>,\left<k_{III}\right>$. 
In the framework of linear elasticity, all these stress intensity factors are linear functions of the loading and can be adimentionalized by $E_f\epsilon\sqrt{h_f}$, where $\epsilon$ is the imposed strain. 
According to the principle of local symmetry, the crack propagates without changing direction if the stress intensity factor in mode~II vanishes. 
In fig.~\ref{Fig:Numeric_K2}, we plot the ratio of shear to opening mode $\left<k_{II}\right>/\left<k_{I}\right>$ as a function of the strip width $W$.
The averaged shear mode $\left<k_{II}\right>$ vanishes for a single value of the width $W_2=15.1 h_f$ for a Poisson ratio $\nu_f=0.25$ (see section~\ref{strip_mat} for other values of $\nu_f$).
As a consequence, only strips with a width $W=W_2$ may thus propagate steadily along a straight path.
Moreover, the change of the sign of $\left<k_{II}\right>$ indicates that $W_2$ corresponds to a stable width.
For a strip width larger than $W_2$, the crack tips indeed undergo a shear loading which would induce a kink of the fracture inward, leading to a decrease of $W$ as the cracks propagate. 
The symmetric situation with a width $W<W_2$ would conversely lead to an increase of $W$. This is an indication that the solution with constant width $W=W_2$ is stable. In any case, the width is thus expected to converge towards $W_2$. The selection of a given width in this problem is, in this sense, comparable to substrate spalling~\cite{Thouless87}, and to more complex geometrical propagation of interacting straight and curved crack~\cite{Xia00}, which have been similarly explained by a vanishing value of $k_{II}$ for a single distance.

In the inset of figure~\ref{Fig:Numeric_K2}, we present  the profiles of stress intensity factors as a function of the position $z/h$ across the film thickness, for the optimal width $W=W_2$.
These profiles indicate that the opening mode ($k_I$) is dominant and roughly constant.
Interestingly $k_{II}$ varies monotonically and only vanishes on average. 
We also note that $k_{III}$ is small but finite (in average it corresponds to 5\% of $k_I$). 
This finite but small value is an indication that the geometry selected (straight front perpendicular to the film) is only a reasonable approximation. 
Since the energy release rate is not constant along the front, some parts will propagate faster than other, and the fracture front should indeed not remain straight. The actual front should slightly tilt sideways (opposite sign of $k_{II}$ on top and bottom faces of the film, and finite $k_{III}$)~\cite{Leblond11}.
 A complete numerical study of this problem would require to find the actual shape of the front in the steady state regime (the one for which $k_{II}=k_{III}=0$ all along the front, with a constant $k_I$).
Such a delicate study would be beyond the scope of the present work.

\subsection{Energetic condition for the propagation of duos: Griffith's criterion.}

In the previous section, we found a steady width $W_2$, which is solely dictated by linear elasticity and thus independent of the loading.
We  now use a steady state argument to determine the conditions for propagation. 
This argument is only relevant for the particular width $W=W_2$, since other widths would lead to unsteady diverging or converging strips. 

Under steady state conditions, the energy released during propagation is the difference between energies far ahead and far behind the fracture front~(fig.~\ref{Fig:Collaborative}b). 
This energy released includes a contribution of the still adhering film, which is identical to the classical case of isolated channel cracks~(fig.~\ref{Fig:Collaborative}a). 
In addition, the elastic energy of the debonded strip is completely released, since the boundaries of the strip are free (fig.~\ref{Fig:Collaborative}b). 
The energy released per unit advance of the front is now given by:
$$
 2 \gamma e h_f  + e W_2 
$$
Wide strips are thus more favorable from an energy release point of view.
However the energy cost for propagation is also higher, as it now involves two fractures across $h_f$ and a  debonding front along $W_2$. 
Propagation is thus energetically possible if 
$$(2 \gamma  h_f  +  W_2) e = 2\Gamma_c h_f + \Gamma_i W_2,$$
where $\Gamma_i$ and $\Gamma_c$ are respectively the debonding and fracture energies of the film.
This relation defines a new condition for propagation of duos of cracks, different from the threshold for isolated cracks~(eq.~\ref{eq:Griffith_isolated}). 
In the case of weak adhesion $\Gamma_i \leq e$,  this condition may be satisfied for sufficiently large widths, even when~eq.~\ref{eq:Griffith_isolated} does not hold. 
In other terms, duos of cracks may propagate in conditions where isolated cracks cannot.

\subsection{Selection of the propagation velocity}
Our experimental estimations of the quasi-static debonding and fracture energies show that both $\Gamma_i(v)$ and $\Gamma_c(v)$ increase with the front velocity (fig.~\ref{fig:properties}d), as it is generally observed in any mode of fracture.
If we take into account velocity dependence, the condition of propagation therefore becomes:
\begin{equation}
 \left(2 \gamma    +  \frac{W_2}{h_f} \right) e \geq 2\Gamma_{c}^0 + \Gamma_i^0  \frac{W_2}{h_f},
 \label{eq:regime_permanent}
\end{equation}
where $\Gamma_i^0$ and $\Gamma_c^0$ respectively correspond to the minimal values of $\Gamma_i$ and $\Gamma_c$, valid for vanishing velocities. Once this relation is satisfied, we expect the excess of residual energy to be dissipated by the fracture and delamination velocity-dependent processes ({\it e.g.} viscous dissipation, diffusion of reacting molecules \cite{Lin07}). 
The resulting quasi-static propagation velocity $v$ is thus set by:
$$ \left(2 \gamma    +  \frac{W_2}{h_f} \right) e = 2\Gamma_{c}(v) + \Gamma_i(v) \frac{W_2}{h_f}. $$ 
The above condition will be compared quantitatively with experimental data in section~\ref{section:sublayer}, where the effect of the finite rigidity of the substrate is studied in details. In the following section we propose an alternative approach based on physical arguments to predict the selection of the width. 

\section{A physical model for the selection of the width}
\label{section:NRJ}

Within the assumption that cracks and delamination fronts are all straight, direct computation of stress intensity factors showed that only strips with a well defined width $W_2$ experience no shear mode, and propagate steadily.
This result is certainly not intuitive since the boundary conditions around each crack are asymmetric conditions.
It is thus difficult to interpret why the shear mode vanishes for a unique width $W_2$. 
We now present an alternative point of view,  where the direction of propagation is deduced from a study of the energy release rate.

Indeed, the ``maximum energy release rate criterion" is also commonly used to predict the direction of fracture propagation in a homogeneous, isotropic material. According to this criterion, fracture propagates in the direction that maximizes the energy release rate for a fixed loading~\cite{cotterell80}. In the case of a smooth crack path (regular in space and in time), in an homogeneous isotropic material, this criterion has been proved to be equivalent to the principle of local symmetry used in the previous section~\cite{Chambolle09, cotterell80}. 
Both criteria nevertheless slightly differ if the path presents a kink~\cite{Hutchinson92, Amestoy92}. 
However, we are here interested in the conditions for a steady propagation, which corresponds to the first case.

\subsection{Elastic energy for a straight strip with width $W$}
\begin{figure}
\centering
 \includegraphics[width=12cm]{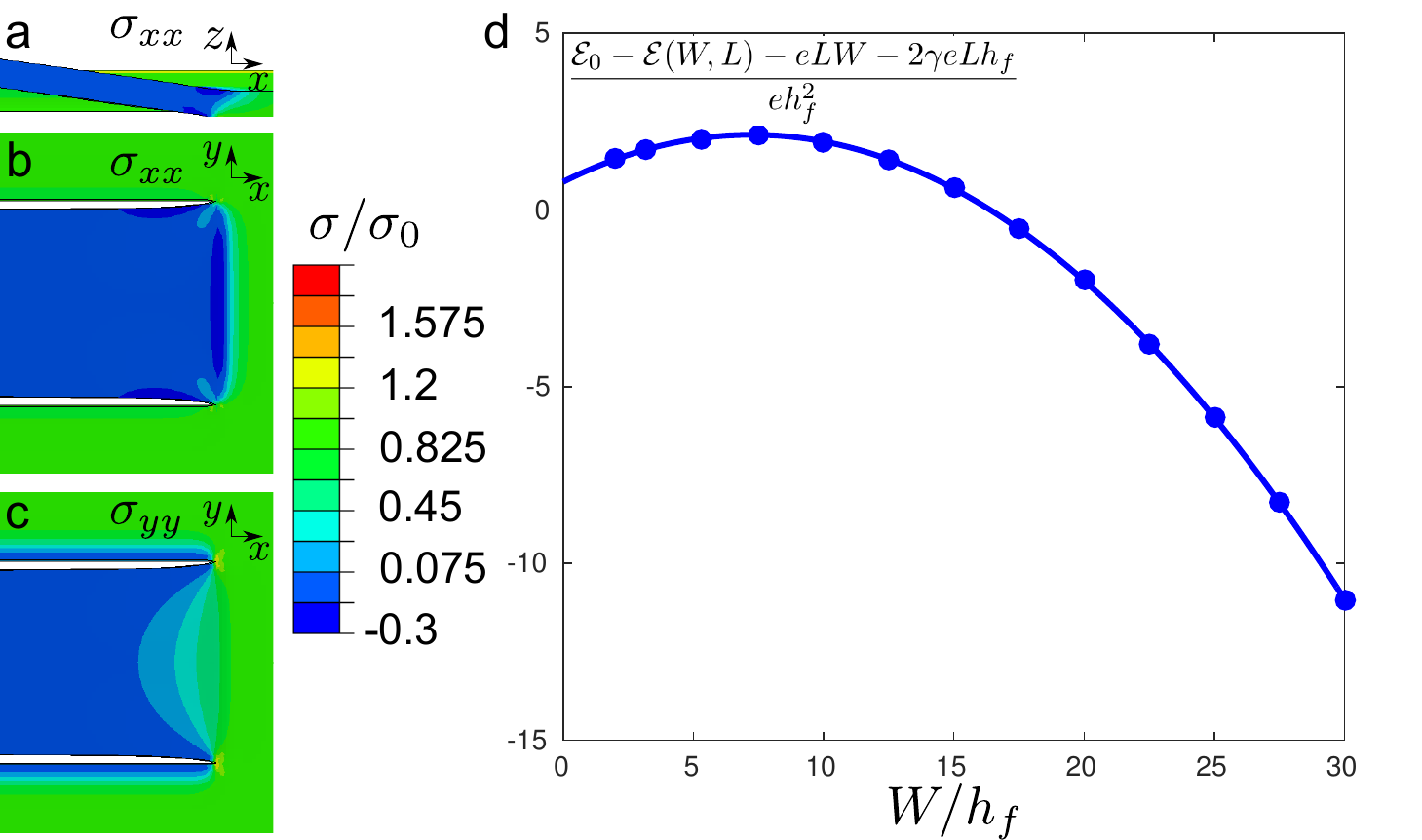}
  \caption{ 
  (a, b, c) Numerical computation of the stress fields. Stress component $\sigma_{xx}$ : the spontaneous tilting of the strip (side cut view of the strip a) releases stresses even ahead of the debonding front (also seen on top view b). Stress component $\sigma_{yy}$ : residual stress is released in the strip except in the vicinity of the delamination front (top view c). 
  (d) Difference in elastic energy due to the delaminated strip ${\cal E}_0 - {\cal E}(W,L) - eLW - 2\gamma e L h_f$ normalized by $eh_f^2$ as a function of  stip width $W/h_f$ for a constant length $L$ (see geometry on figure~\ref{Fig:Collaborative}b). The plot presents a minimum and can be fitted with a quadratic function $-\alpha W^2+\beta h_f W$ with $\alpha \simeq 0.0254$,$\beta \simeq 0.368$}
  \label{fig:EvsW}
\end{figure}

Similarly to the previous section, we consider a film under bi-axial tensile stress covering an infinitely rigid substrate.
We will use the initial energy of this film as a reference ${\cal E}_0$ and we propose to estimate the elastic energy of the system ${\cal E} (W,L)$ as a long strip of length $L$ and width $W$ ($L\gg W \gg h_f$) delaminates from the substrate (fig.~\ref{Fig:Collaborative}b).
As suggested by the experiments, the delamination front is assumed to be straight. We use the coordinates $x,y$ in the plane of the coating, $y$ being oriented along the delamination front.
The previous finite elements calculations provide the stress field distribution on the film displayed in fig.~\ref{fig:EvsW}. 
The observation of the stress field shows that residual stresses are decreased in two independent regions: the delaminated strip and a pair of narrow bands of the adhering film along the fracture path. 
We thus split the difference in elastic energy, ${\cal E}_0-{\cal E} (W,L) = {\cal E}_{\rm strip}(W,L)+ {\cal E}_{\rm cut}(L)$, into the contributions of these  respective regions.
The term ${\cal E}_{\rm cut}(L)$ is similar to the case of classical isolated cracks and does not depend on the width of the strip.

We have seen that far enough from the delamination front, the steady state difference in elastic energy  is given by $eW+ 2 \gamma e h_f $ per unit strip length (eq.~\ref{eq:regime_permanent}).
Integrating this difference over the length of the strip would thus lead to ${\cal E}_{\rm cut}(L)= 2 \gamma e h_f L$ and $ {\cal E}_{\rm strip}(W,L)= e{\cal A}$, where ${\cal A}$ is the debonded area. 
However, this expression neglects boundary effects close to the crack tip and to the delamination front. 
Indeed, the elastic field along the edges of the cut is not invariant along $x$ (steady state elastic field) close to the fracture tip. 
The steady state energy for the cut  is only valid a few thickness away from the tip and should be corrected into ${\cal E}_{\rm cut}(L)= 2\gamma e h_f ( L + \zeta h_f ),$  where $\zeta$ is a numerical factor.
In a similar manner, the delamination front perturbs the complete relaxation of the residual strain in the strip.
We computed numerically the energy of the finite size system $ {\cal E}(W,L)$ using the numerical scheme developed in section~\ref{duosKII}. 
We plot in fig.~\ref{fig:EvsW}a the elastic energy $ {\cal E}_0 - {\cal E}(W,L) - eLW - 2\gamma e L h_f =  {\cal E}_{\rm strip}(W,L)-e{\cal A}  + 
2\gamma \zeta e h_f^2 $.
The strip length $L$ was kept constant (as well as ${\cal E}_{\rm cut}$), and only the width $W$ of the strip was varied.
In other words, this graph represents the evolution of 
$eh_f^2{\cal G} (W) = ({\cal E}_{\rm strip}(W,L)-e{\cal A} )$ shifted by a constant  value.
An essential feature of the additional term in energy release ${\cal G}(W)$ is to present a maximum.
In the following section, we provide a physical interpretation of this non-monotonic variation as the result of two elastic perturbations that take place close to the delamination front. We  show in section~\ref{W_2NRJ} how this property determines the crack path.

\subsubsection{Strain along the width at the boundary:  ``a surface effect"}
The boundary condition at the end of the strip in  contact with the attached film prevents the residual strain to be completely relaxed in the $y$ direction (fig.~\ref{fig:EvsW}c).
In a first approximation, the problem is thus similar to a planar strip with three free edges but under a tensile strain $\epsilon_{yy}= \epsilon_0$ along the last one.
Due to the laplacian nature of elastic equations for plane stress, the boundary condition has an influence over a characteristic length on the order of $W$. As a consequence, the region of the strip remaining under strain has an area proportional to $W^2$, leading to an elastic energy $\alpha e W^2$, where $\alpha$ is a numerical prefactor depending only on the Poisson ratio of the film $\nu_f$.
We thus expect a correction of our estimation of the released energy with a term $-\alpha e W^2$. 
As a first approximation, this correction should only depend on $\nu_f$ and be independent of any other material parameter, even if the substrate is not considered as infinitely rigid (see section~\ref{section:sublayer}).

We note however that this correction depends on the geometry of the delamination front. In this paper we assume that the front is straight, based on optical observation (see for example figures~\ref{Fig:W_vs_h}a to c). This assumption remains to be justified by a determination of the geometry of the delamination front using interfacial fracture mechanics, a challenging question that goes beyond the scope of this paper.

\subsubsection{Tilting the strip: a line effect}\label{line}
In contrast with $\sigma_{yy}$ stresses, $\sigma_{xx}$ residual stresses are completely relaxed in the strip.
However more energy can be released in the remaining coating if the strip is tilted  (fig.~\ref{fig:EvsW}a and b). 

In experiments, this out-of-plane displacement is evidenced through the interference fringes usually visible in the debonded area (fig.~\ref{Fig:W_vs_h}). It is also visible in the finite element simulation presented in figure~\ref{fig:EvsW}a where the strip remains straight in the debonded zone. Additional effects such as gravity, electrostatic forces, or surrounding airflow usually lead to a re-attachment of very long strips after propagation has taken place, so that tilting is not evidenced in figure~\ref{fig:properties}b. In our simplified model of the delamination front, we do not consider these potential additional bending forces.

Tilting the strip provides some localized shear across the thickness of the film in the vicinity of the delamination front. 
This situation is very similar to the boundary condition at the free edges of usual (isolated) channel cracks, where the shear in the thickness relaxes stresses over a distance of order $h_f$ along the crack path.
This additional contribution leads to a correction of the estimation of the released energy by a term $\beta e W h_f$, where $\beta$ is a numerical factor.  
As a first approximation, we expect $\beta$ to mainly depend on $\nu_f$ if the substrate is infinitely rigid. 
We study in section~\ref{section:quantitative} the effect of a substrate of finite stiffness. \\

\subsubsection{Combining both effects}
Implementing these two contributions in the strip leads to a correction of the the estimation of the energy released:
\begin{equation}
{\cal G}(W) = ({\cal E}_{\rm strip}(W,L)-e{\cal A})/eh_f^2 = \beta (W/h_f) - \alpha (W/h_f)^2.
\label{eq:GW}
\end{equation}
which is in agreement with the numerical results (fig.~\ref{fig:EvsW}d).
Indeed, a quadratic function fits fairly well the difference ${\cal E}-eWL$ and gives $\alpha \simeq 0.0254$ and $\beta \simeq 0.368$ for a thin film with a Poisson ratio $\nu_f=0.25$ deposited on a rigid substrate. 
We discuss in section~\ref{section:sublayer} the dependence of $\alpha$ and $\beta$ with the Poisson ratios and the finite stiffness of the substrate.
The elastic field close to the delamination front is therefore well captured by these two ingredients: a line effect favoring large widths (linear term in $W$) and a surface effect penalizing large widths (quadratic in $W$).

It might be intuitive to deduce from this result that the system selects an optimal width which maximizes ${\cal G}$. 
However, such an optimization is not grounded and would lead to a width $W_2\simeq 7 h_f$, which is half the prediction from stress intensity factors.
The role of ${\cal G}$ is nevertheless crucial in the selection of the width as the fracture propagates. 
We discuss this selection in the following section.

\subsection{Selection of the width} 
\label{W_2NRJ}
\begin{figure}
\centering
 \includegraphics[width=12cm]{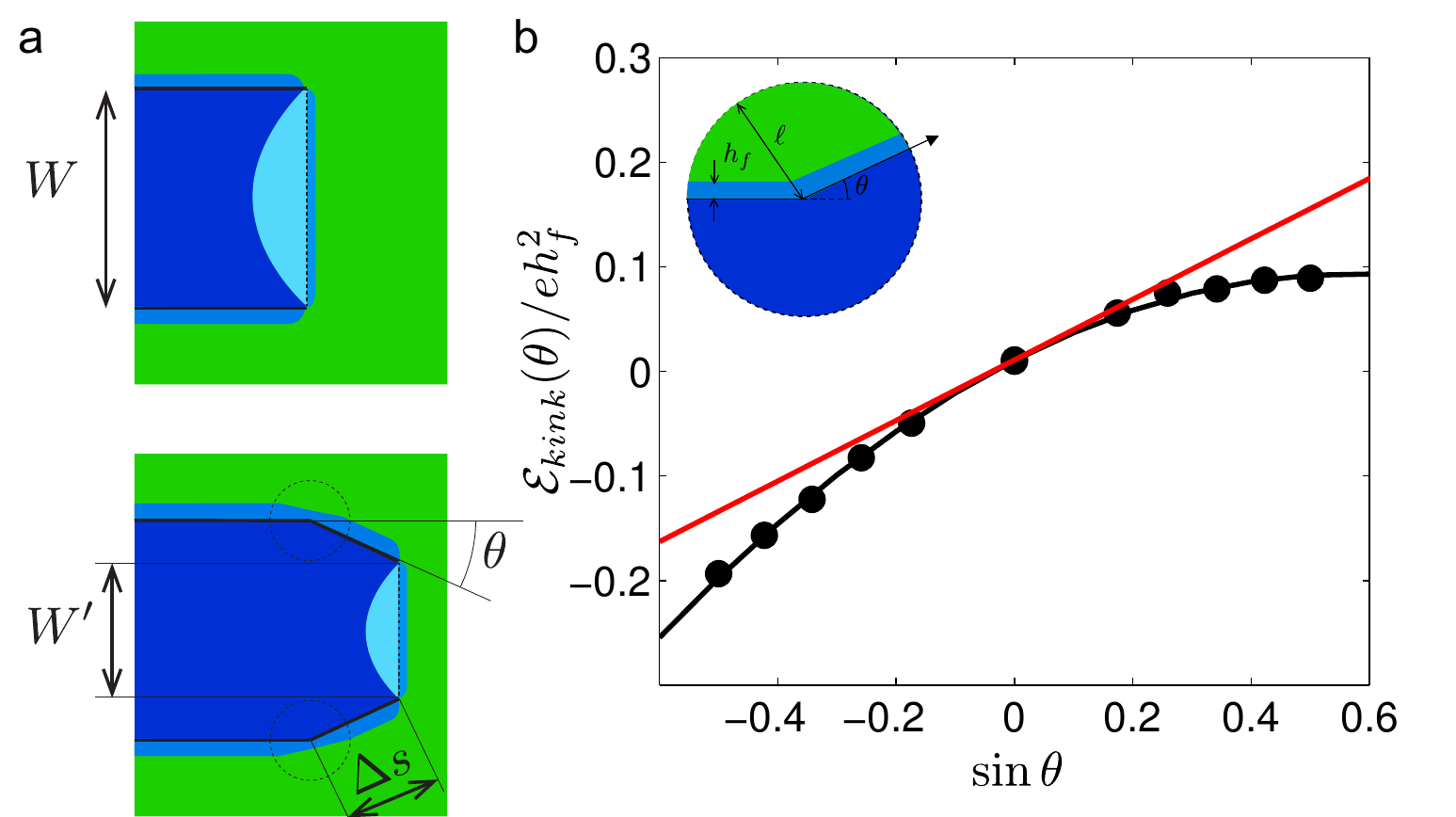}
   \caption{(a) Starting from a straight strip (top) we consider the propagation of a fracture  in a direction $\theta<0$ by a distance $\Delta s$. 
In the sketches, the color symbolizes qualitatively residual elastic energy (as in figures 4 and 6).
(b) Additional adimensionalized released energy due to the kink  $\mathcal{E}_{\rm kink}(\theta)/ eh_f^2$  . (Inset) Zoom into a kink of angle $\theta$ in the crack path.}
  \label{Fig:Courbure}
\end{figure}

Starting from an initial long strip of width $W$ and length $L$ ($L\gg W$), we propose to determine in which direction the symmetric cracks should propagate. 
To address this question, we compute the energy released by propagation as a function of the direction of propagation.
Following the previous sections, the elastic energy in this initial state is given by:
$${\cal E} = {\cal E}_0 - e{\cal A}  - e{\cal G} (W) - 2  \gamma e h_f (L + \zeta h_f ).$$
We now consider the symmetric propagation of a pair of cracks forming an angle $\theta$ with the axis of symmetry (fig.~\ref{Fig:Courbure}a). 
If both cracks advance by a distance $\Delta s$, the delamination front is also translated by a distance $\Delta x=\Delta s \sin\theta.$ 
Although the kinked configuration is more complex than the previous uniform strip, we will assume that the expression for the elastic energy in the strip remains valid.
In other words, we assume ${\cal E}'_{\rm strip} =  e {\cal A}' + e{\cal G} (W')$, where 
$\cal A'$ is the new area of the delaminated part, and $W'$ the width of the front.

However, the elastic stress along the cut is altered by the kink and the expression of the energy release is more subtle than ${\cal E}'_{\rm cut} = 2  \gamma e h_f (L'+ \zeta h_f )$, where $L'=L+\Delta s $ is the new crack path length. We computed numerically the stress field in the vicinity of a kink of angle  $\theta$ (fig.~\ref{Fig:Courbure}b).
We consider the elastic energy $\mathcal{E}_{\rm sector}(\theta,l)$ in a disk of radius centered on the kinking point (inset of figure~\ref{Fig:Courbure}b).
As a rough approximation, this energy would be simply $e l^2(\pi-\theta) - \gamma e h_f 2l$.
However, the presence of the kink modifies significantly $\mathcal{E}_{\rm sector}(\theta,l)$. 
We represent in figure~\ref{Fig:Courbure}b the  correction of the released energy $\mathcal{E}_{\rm kink}=e l^2(\pi-\theta) - \mathcal{E}_{\rm sector} - 2\gamma e h_f l $ as a function of $\theta$.
This correction is independent from the radius of the disk as long as the condition $l\gg  h_f$ is verified.
We obtain a positive correction if the kink angle is positive (more energy is released), which promotes the outward propagation of the strip.
As a first order approximation, this correction varies linearly with $\sin\theta$, leading to the contribution of the kink  $\mathcal{E}_{\rm kink} =  e \delta \gamma h_f^2 \sin\theta$, with $\delta = 0.29$. Note that $\mathcal{E}_{\rm kink}$ is not exactly an odd function of $\theta$ that would vanish in the case of isolated channel cracks as both the contributions of the complementary sectors are added.
Additional non-linear terms lead to a slight decrease of the released energy if a kink is formed, which finally favors the selection of straight paths in standard channel cracks.\\ 

If we add the different contributions, the energy obtained after the propagation of the cracks is given by 
${\cal E}' = {\cal E}_0 - {\cal E}'_{\rm strip} - {\cal E}'_{\rm cut} - 2 \mathcal{E}_{\rm kink}$.
We finally obtain the elastic energy released during propagation by a distance $\Delta s$ in a direction $\theta$ as the difference:
$$ \mathcal{E}_r = \mathcal{E}- \mathcal{E}'= e [{\cal A' -A}] + e[{\cal G}(W') - {\cal G}(W) ] + 2 \gamma e h \Delta s + 2\mathcal{E}_{\rm kink}.$$
under the condition $\Delta s \gg \gamma h_f$.
We are however interested in the energy release rate $ (\mathcal{E}- \mathcal{E}')/\Delta s$ as the propagation distance $\Delta s$ vanishes. 
Although $\mathcal{E}_{\rm kink}$ cannot be easily computed, its value is constant for $\Delta s \gg  h_f$ and vanishes for $\Delta s=0$ as the kink vanishes as well.
As a first approximation, we assume that $\mathcal{E}_{\rm kink}$ reaches its asymptotic value as the crack propagates by a distance on the order of a thickness $ h_f$. We thus estimate the derivative as $ \mathcal{E}_{\rm kink}/\Delta s \sim  \delta \gamma  e h_f \sin\theta $, when $\Delta s$ vanishes.\\

We finally deduce the energy released rate for the propagation of cracks  in the direction~$\theta$: 
$$  \frac{d{\cal E}_r}{ds} = 
e  \frac{d{\cal A}}{ds} - e  \frac{d{\cal G} (W) }{dW} \frac{dW}{ds}    +  2 \gamma eh_f   + 2e  h_f { \delta \gamma}\sin\theta. 
$$
If we input in this last equation the geometrical relations $dA=W ds\cos\theta$ and $dW=2ds\sin\theta$ and the expression for ${\cal G }(W)$ determined in eq.(\ref{eq:GW}), we obtain:
\begin{equation}
 \frac{d{\cal E}_r}{ds} = e [ W\cos\theta + 2 \gamma  h_f -2(2\alpha W - \beta' h_f)\sin\theta ],
\label{eq:released}
\end{equation}
with $\beta'= \beta + \delta \gamma$. 
Following Griffith criterion, the cracks propagate simultaneously with the delamination front if this elastic energy release rate balances the fracture energy of the two cracks $2 \Gamma_c h_f$ plus the delamination energy $\Gamma_i W\cos\theta$. 
The condition for propagation is therefore:
\begin{equation}
 e [ W\cos\theta + 2 \gamma  h_f -2(2\alpha W - \beta' h_f)\sin\theta ] =\Gamma_i W\cos\theta + 2\Gamma_c h_f
\label{eq:griffith}
\end{equation}
Note that we only need to determine the direction of propagation of the fracture in the film, since the the interfacial delamination front is forced  to follow the weak interface.
In this case, we use a general version of the criterion of maximum energy release rate: the direction of propagation is the one that maximizes the difference of elastic energy release rate with the dissipation rate~\cite{Chambolle09,cotterell80}.
The direction of propagation is thus set by 
$\partial (d{\mathcal{E}_r}\,/ds - \Gamma_i W\cos\theta - 2\Gamma_c h_f) / \partial \theta = 0$, which leads to:
 \begin{equation}
e W\sin\theta + 2e(2\alpha W -\beta' h_f)\cos\theta = \Gamma_i W \sin\theta
\label{eq:release}
\end{equation}
If we combine equations~(\ref{eq:griffith}) and (\ref{eq:release}), we finally obtain :
\begin{eqnarray}
 (e-\Gamma_i) W &=& 2( \Gamma_c - \gamma e) h_f \cos\theta   \label{eq:Griffith} \\
 \sin\theta &=& -\frac{e}{\Gamma_c-\gamma e }(2\alpha W/h_f - \beta') 
\label{eq:angle}
\end{eqnarray}
This last relation shows that the propagation is straight ($\theta=0$) only for a given distance width $W=W_2$, where
\begin{equation}
W_{2} = \frac{\beta'}{2\alpha}h_f,  \label{eq:W2}
\end{equation}
This width is stable as the sign of $\theta$ compensates the deviation from $W_2$.
Indeed, if $W$ is larger than $W_2$, the propagation angle is negative and both cracks move inwards. Conversely, if $W$ is smaller than $W_2$, the propagation angle is positive and the two fractures move outwards. Since both prefactors $\alpha$ and $\beta'$ are only set by elasticity, independently of the fracture energy and the loading of the film, the expected width should be robust, in agreement with our experimental observations
The parameter $\delta$ is difficult to compute exactly and we arbitrary take $\beta'=0.76$, which corresponds to the width $W_2/h_f\sim 15$ obtained in the numerical study of the stress intensity factor. 
This value corresponds to $ \delta\gamma = 0.4$ which is compatible with our order of magnitude estimate $\delta \gamma \sim 0.18$.

\subsection{Convergence length}
In the previous section, we have shown that the selected width $W_2$ was stable, which leads to uniform strips. 
Indeed, if the initial width of the strip $W$ differs from $W_2$, the eq.~\ref{eq:angle} predicts a convergence of $W$ towards $W_2$.  
Such a rapid convergence towards the optimal width is a robust characteristic observed in the experiments (fig.~\ref{Fig:W_vs_h}b and c).
When the fracture is deviated by a defect in the film, it effectively quickly goes back to the optimal width.
We wish here to estimate the characteristic length scale of this convergence.

We assume that the energy estimates in (\ref{eq:GW})  are valid for strips with slowly varying width, so that equations (\ref{eq:angle}) and (\ref{eq:Griffith}) are still valid. 
If we introduce $\sin \theta = \frac{1}{2} dW/ds$ in Eq.~(\ref{eq:angle}), we obtain: 
\begin{equation}
 \frac{dW}{ds}=\frac{1}{\xi}(W_2-W),
\end{equation}
which leads to an exponential convergence from $W$ to $W_2$ over the characteristic length $\xi$ given by:
\begin{equation}
 \xi=\frac{\Gamma_c-\gamma e }{4 e \alpha}h_f.
\end{equation}
 As $(\Gamma_c-\gamma e)/2e$ is of the order of unity in our experiments, as well as the parameter $\beta'$, which leads to $\xi \sim W_2$. Experimentally, we effectively observed typical convergence lengths of the order of $W_2$ (fig.~\ref{Fig:W_vs_h}b and c).\\

As a conclusion, although our analytical approach is approximate, it provides a physical interpretation for the robust selection of a particular strip width $W_2$.
Three main physical ingredients indeed dictate the value of $W_2$, the surface and line effects due to the boundary layer at the attached end of the strip and the effect of a curvature of the crack path.
Moreover this analytical approach predicts how $W_2$ as a function of elastic properties of the film. 
In the next section we will focus on the case of a substrate of finite stiffness and show how the analytical approach is also relevant in this situation.

\section{Influence of the Poisson coefficient and of the finite rigidity of the substrate for a quantitative comparison with experiments}
\label{section:sublayer}

In the previous sections, the substrate was assumed to be infinitely rigid. 
However, thin sublayers were used in the experiments to modify adhesion properties.
In order to compare quantitatively our experimental data with our theoretical predictions, we study in this section the influence of the substrate rigidity and more generally of the material properties of the coating in the selection of the optimal width.  
This quantitative study is important to determine the conditions of stability of the coatings.

\subsection{Influence of the Poisson ratio of the coating}
\label{strip_mat}
In the case of an infinitely rigid substrate, the selected width $W_2$ was found to be independent of the Young modulus of the coating.
However $W_2$ is a function of the Poisson ratio and we investigate this dependence in the following section.

\begin{figure}
\centering
  \includegraphics[width=12cm]{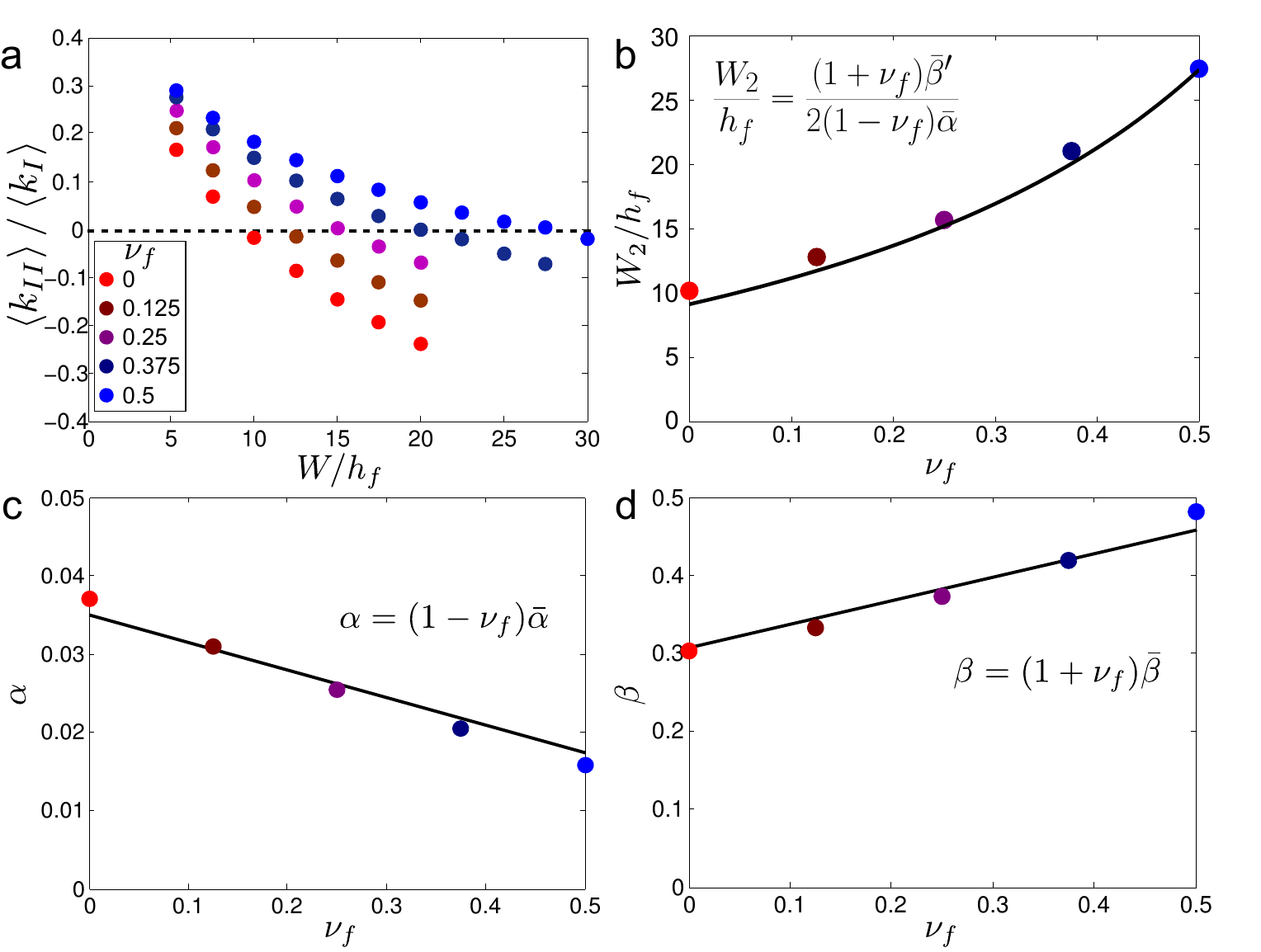}
  \caption{(a)  Numerical computation of the average shear loading $\left<k_{II}\right>/\left<k_I\right>$ along fracture front of a delaminating strip, normalized by opening mode, as a function of strip width $W$, for different Poisson coefficient of the thin film ($\nu_f=0, 0.125, 0.25, 0.375, 0.5$). (b) The corresponding equilibrium strip width $W_2$ (for which $k_{II}=0$) slightly increases with $\nu_f$, as captured by the 
 the prediction of the physical model in black continuous line ($W_2/h_f=(1+\nu_f)\bar\beta'/2(1-\nu_f)\bar\alpha$) with $\bar \beta'=\bar\beta + 0.32$.  (c,d) Numerical computation of parameters  $\alpha$ and $\beta$ for our physical model as a function of $\nu_f$, compared to (black continuous line) prediction $\alpha=(1-\nu_f)\bar \alpha$. (c) Numerical computation of parameter $\beta$ compares well with theoretical prediction in black line $\beta=(1+\nu_f)\bar \beta$ (D).}
  \label{Fig:Poisson}
\end{figure}

We first extend our numerical study of stress intensity factors developed in section~\ref{duosKII}  for a coating with a Poisson coefficient $\nu_f$ between 0 and 0.5.
We plot in fig.~\ref{Fig:Poisson}a the corresponding ratio $\left<k_{II}\right>$ as a function of the adimensionalized width $W/h_f$.
In the whole range of Poisson coefficient, $k_{II}$ vanishes for a given width $W_2$, which corresponds to a stable self-peeling strip. 
We find that $W_2$ slightly increases with $\nu_f$ ranging from $W_2=10h_f$ for $\nu_f=0$ to $W_2=25h_f$ for $\nu_f=0.5$ (fig.~\ref{Fig:Poisson}b).\\

Our physical model  constitutes a useful  tool to interpret  this dependence of $W_2$ with $\nu_f$. 
Indeed, following equation~(\ref{eq:W2}), we only need to determine the evolution of the numerical factors $\alpha(\nu_f)$ and $\beta'(\nu_f)$ with the Poisson coefficient.
The layer is initially subjected to equibiaxial residual stress $\sigma_0$, which results into a density of elastic energy per unit area $e= h_f\sigma_0^2(1-\nu_f)/E_f$.
If stresses in one direction are relaxed ({\it e.g.} $\sigma_{yy}=0$) while strains in the other direction are maintained ({\it e.g.} $\epsilon_{xx}=\epsilon_0$), the energy per unit surface becomes $h_f \sigma_0^2(1-\nu_f)^2/2E_f= e (1-\nu_f)/2$.
This simple results provides a prediction for the variation of the coefficients ($\alpha,\beta,\gamma$) with $\nu_f$.
Indeed, the delaminated strip relaxes freely in $x$ and $y$ directions, excepted in the vicinity of the delamination where strains in the $y$ direction are not relaxed (fig.~\ref{fig:EvsW}c).
In this region the remaining energy density (or in other words, the deficit in elastic energy released relative to a free strip) is thus proportional to $e (1-\nu_f)/2$ , which leads to a dependence of $\alpha$ with $\nu_f$:
\begin{equation}
 \alpha = (1-\nu_f)\bar\alpha,
 \label{eq:alpha}
 \end{equation}
where $\bar\alpha$ is an universal constant independent of material properties. 
The same situation occurs in the region of the bonded film in the vicinity of the debonding front (stresses are partially relaxed in the $x$ direction by shear, but not in the $y$ direction).
The density of elastic energy released by tilting the strip is thus proportional to the difference $ e - e(1-\nu_f)/2$, which now leads to a dependence of $\beta$ with $\nu_f$ in the form
\begin{equation}
\beta=(1+\nu_f)\bar \beta,
\label{eq:beta}
 \end{equation} 
where $\bar\beta$ is also a universal constant. 
The same argument applied in the region of the kink finally leads to a similar result
$\gamma=(1+ \nu_f) \bar \gamma,$
where $\bar\gamma$ is again a universal constant. 
If we combine  both effects, we finally obtain for the coefficient $\beta'$:
\begin{equation}
 \beta'=(1+\nu_f)\bar \beta'
\label{eq:betaprime}
 \end{equation}
 Both coefficients $\alpha$ and $\beta$ are determined by fitting the elastic energy of the strip $({\cal E}_{\rm strip}(W,L)-e{\cal A})/eh_f^2$ obtained through finite element simulations with a quadratic function $-\alpha (W/h_f)^2 + \beta (W/h_f)+c$.
We plot in fig.~\ref{Fig:Poisson}c and \ref{Fig:Poisson}d the variation of $\alpha$ and $\beta$ for different values of  $\nu_f$ ranging from 0 to 0.5. 
This evolution is in good agreement with eqs.~(\ref{eq:alpha}) and (\ref{eq:beta}), from which we extract the constants $\bar\alpha=0.0349$, $\bar\beta=0.306$.
Introducing eqs.~(\ref{eq:alpha}) and (\ref{eq:betaprime}) into eq.~(\ref{eq:W2}), finally leads to the dependence of the selected width with the Poisson coefficient:
\begin{equation}
 \frac{W_2}{h_f}=\frac{(1+\nu_f)\bar \beta'}{2(1-\nu_f)\bar\alpha}.
\end{equation}
This relation is in good agreement with the evolution of the width determined directly from the computation of stress intensity factors. 
In this fit we have corrected $\bar\beta' =  \bar \beta + 0.32$ using the value previously determined in section~\ref{strip_mat} with $\delta \bar \gamma = \delta \gamma/(1+\nu_f)$. As a conclusion, although our physical approach is approximate, it brings quantitative prediction of the influence of the Poisson coefficient on the selected width.
In this section the substrate was still considered as infinitely rigid. In the following section we study influence of the finite rigidity of the substrate.

\subsection{Influence of a sublayer and quantitative comparison with experiments }
\label{section:quantitative}
We now consider the effect of the finite stiffness of the substrate.
In actual experiments we used a sublayer of silicate in order to tune the adhesion properties of the coating.
We define as $h_{sc}$, $E_{sc}$ and $\nu_{sc}$ the thickness, the Young modulus and the Poisson coefficient of the sublayer, respectively.
Note that in the limit of thick sublayer, the effect of the underlying substrate is screened, and the problem is equivalent to a film deposited on an elastic substrate. We assume that the presence of the sublayer does not change qualitatively the physics of the selection of the width, but modifies the quantitative value of $W_2$. In terms of independent parameters, the width selected in the duos of crack (eq.~\ref{eq:GW}) will now depend on the ratio of Young moduli of the film and the sublayer, the Poisson coefficients of the film and the sublayer and the ratio of the thickness of the film and the sublayer.
Dimensional analysis shows that this dependence can be rewritten into $W_2/h_f={\cal F}(E_{sc}/E_f,\nu_f,\nu_{sc},h_{sc}/h_{f})$.
In the following sections, we first predict the selection of the width selection directly using the principle of local symmetry on stress intensity factors determined numerically. 
We then interpret these results with our physical model, before comparing quantitatively these theoretical predictions with our experiments.

\subsubsection{Influence of the rigidity and the thickness of a sublayer}
We first  conduct finite element calculations for fixed Poisson coefficients $\nu_f=\nu_{sc}=0.25$. We determine the equilibrium width from the estimation of stress intensity factors and the principle of local symmetry.
We  present in fig.~\ref{Fig:Young_contrainte}a  $W_2/h_f$ as a function of the thickness ratio $h_{sc}/h_{f}$, for different  ratios of Young moduli $E_{sc}/E_{f}$.
Note that $h_{sc}=0$ corresponds to the calculation presented in section~\ref{duosKII} for a thin film directly coated on a rigid substrate, where we found $W_2/h_f =15.1$. The different curves thus start from this reference value.
Coating the substrate with a sublayer more rigid that the film almost does not affect the selection of the width. 
Nevertheless, $W_2$ increases significantly with with $h_{sc}$ if the sublayer is less rigid than the film. 
More generally, increasing the compliance of the sublayer results into wider strips. 
This variation eventually reaches a plateau value for a typical thickness ratio that also increases as the ratio $E_{sc}/E_{f}$ decreases.

\subsubsection{Interpretation with the physical model}
\begin{figure}
\centering
  \includegraphics[width=12cm]{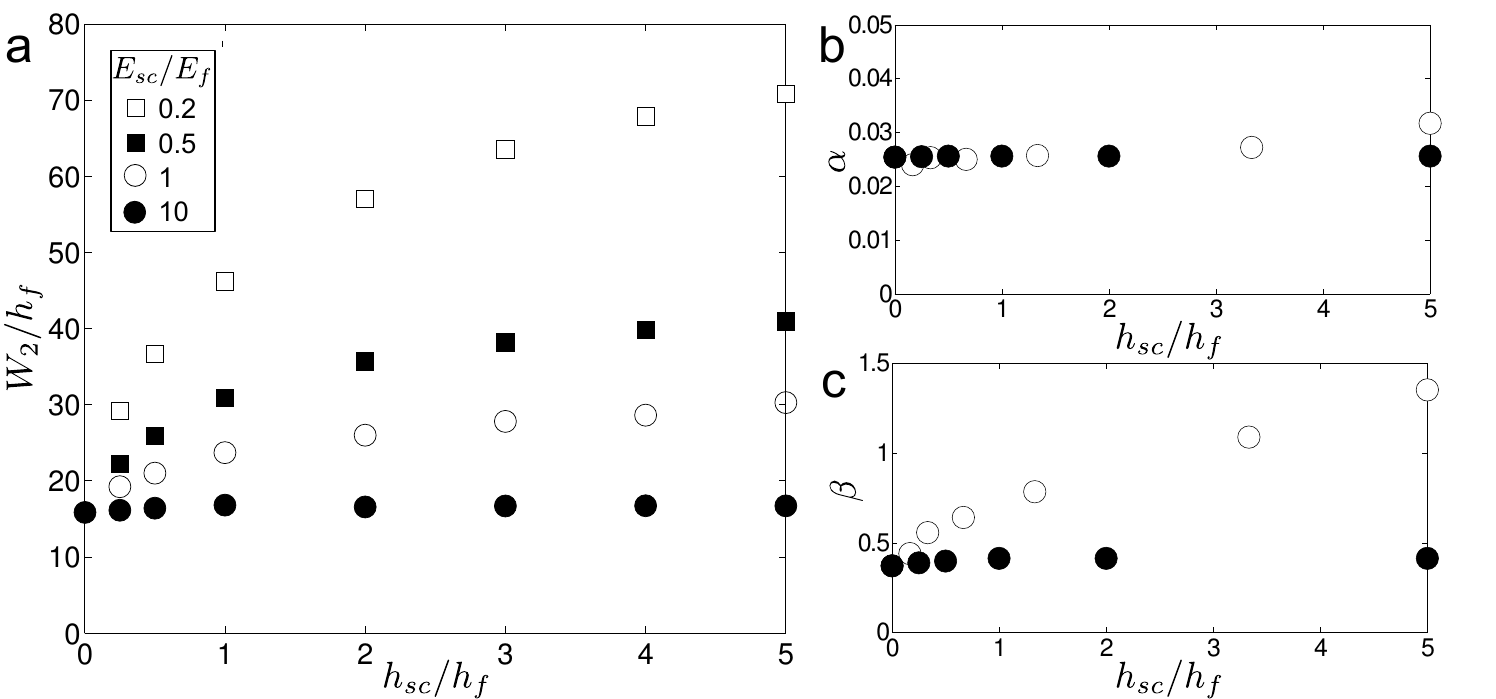}
  \caption{(a) The optimal width (predicted through the numerical calculation of stress intensity factors) as a function of the sublayer to layer thickness ratio $h_{sc}/h_f$, for different rigidity ratio $E_{sc}/E_f$. (b) and (c) Respective evolution of the coefficient $\alpha$ and $\beta$ as a function of the sublayer thickness for $E_{sc}/E_f=10$ and $1$.
}
  \label{Fig:Young_contrainte}
\end{figure}

We emphasize that the phenomenological model described in section~\ref{section:NRJ} provides physical understanding on the numerical calculation of stress intensity factors. 
We indeed found that the steady width mainly results from a balance between a surface and a line effect in the vicinity of the delamination front.
Both effects depend differently on the properties of the sublayer.
We expect the distribution of stresses in the delaminated strip to be independent of the elastic properties of the foundation.
This independence is confirmed in fig.~\ref{Fig:Young_contrainte}b where the calculated parameter $\alpha$ is shown to be almost independent of  thicknesses ratio $h_{sc}/h_{f}$ and Young moduli ratio ($E_{sc}/E_f=1$ and 10).

Conversely, we expect the coefficient $\beta$ to strongly depend  on the properties of the elastic foundation. 
We indeed discussed in section~\ref{line} the similitude of the line effect (resulting from the tilting of the strip) with the energy released by the propagation of a standard isolated channel crack.
In this last case, the energy release is characterized by the parameter $\gamma$ whose dependence on material properties of the substrate has been thoroughly studied by Beuth~\cite{Beuth92}.
We plot in fig.~\ref{Fig:Young_contrainte}c the variation of $\beta$ with the ratio $h_{sc}/h_{f}$ for to ratios of Young moduli, which confirms this dependence : larger release is obtained on more compliant substrate. 
A constant parameter $\alpha$ together with $\beta$ increasing with substrate compliance finally explains why the strip width $W_2/h_f$ (predicted as $\beta/2\alpha$) in fig.~\ref{Fig:Young_contrainte}a is larger on more compliant and on deeper sublayer. 

\subsubsection{Quantitative comparison with experiments}
\begin{figure}
\centering
  \includegraphics[width=12cm]{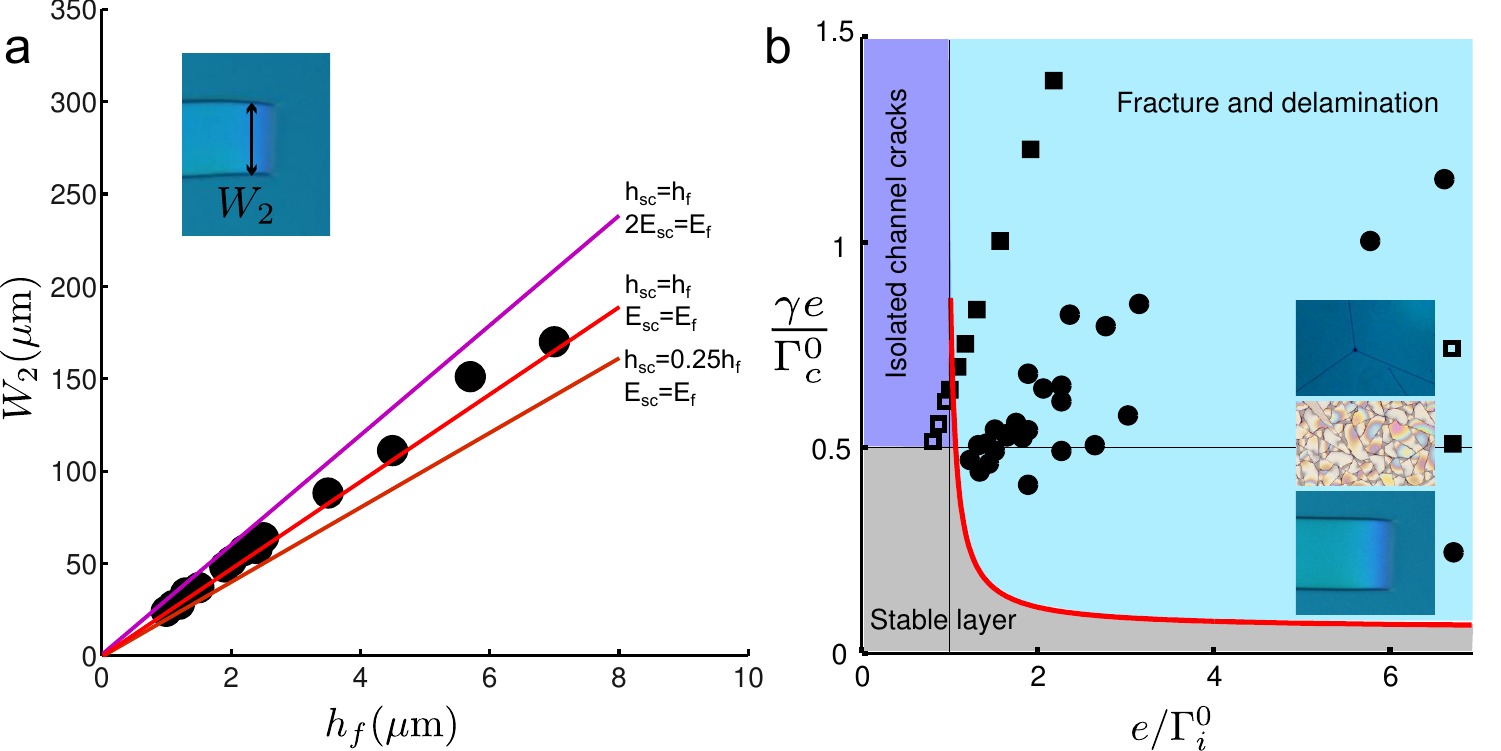}
  \caption{(a) Comparison of the steady strip width $W_2$ computed numerically (colored lines for different conditions) with our experimental data. 
 (b) New stability condition of for a thin coating in the plane $(\gamma e/\Gamma_c^0, e/\Gamma_i^0)$, where $e$ is the available residual elastic energy in the coating per unit surface,
  $\Gamma_c^0$ and $\Gamma_i^0$ the fracture and adhesion energy, and $\gamma$ a numerical prefactor that depends on the mismatch of material properties between the substrate and the coating.
 The $x$-axis corresponds to the comparison of the residual energy density to the interfacial energy, whereas the $y$-axis compares residual elasticity to the material toughness. Isolated channel cracks occur above the line $(2\gamma e/\Gamma_c^0= 1)$,  delamination is energetically possible on the right of vertical line $e/\Gamma_i^0=1$, but only between free edges which require preliminary fractures. The propagation of crack duos (for $E_f=E_{sc}$, $\nu_f=0.25$ and $h_f=h_{sc}$) is possible above the red curve. Open squares represent experimental observation of channel crack without delamination, filled squares with delamination, filled circles correspond to the delaminated strip studied here.}
  \label{Fig:DiagPhase}
\end{figure}

In our experiments the adhesion energy was tuned by depositing a sublayer on the surface of the substrate (this sublayer was thin enough to remain stable with respect to fracture). 
We performed experiments with different values of $h_f$ to probe the linear relation between the width of the strip and its thickness that was predicted theoretically (eq.~\ref{eq:W2}). 
The experimental data are fairly well described by a linear relation $W_2=25h_f$ (fig.~\ref{Fig:W_vs_h}d).
However, the thickness of the sublayer was held close to a fixed value ($h_{sc}=1~\mu$m) and the ratio $h_{sc}/h_f$ (ranging from 0.15 to 2), was not constant for these experiments. 
Moreover, the elastic properties of the layer and sublayer slightly differ for the different coatings ($E_{sc}/E_f$ ranging from 0.5 to 10). As a consequence, we do not expect the ratio  $W_2/h_f$ to be rigorously constant. We thus compare the experimental data to theoretical predictions for two extreme cases ($h_{sc}=h_{f}$, $2E_{sc}=E_f$) and ($h_{sc}=0.25h_{f}$, $E_{sc}=E_f$). These numerical predictions and in particular the intermediate case ($h_{sc}=h_{f}$, $E_{sc}=E_f$) are consistent with the experimental measurements (fig.~\ref{Fig:DiagPhase}a). We finally observe that the adhesion energy (represented in color scale) does not modify the width of the strip, as predicted in our model.

\subsection{Stability diagram}
In the standard description of thin films two modes of failure were classically presented. 
Straight isolated channel cracks can indeed propagate if the energy release rate $2 \gamma e$ exceeds the fracture energy $\Gamma_c^0$.
They for instance lead to the hierarchical pattern observed in old paintings.
Delamination can also occur when the energy release rate $e$ exceeds the adhesion energy $\Gamma_i^0$.
However, delamination requires free edges to initiate. 
In practice, this failure mode usually follows the propagation of channel cracks~\cite{pauchard06, Lazarus11}. Both modes are described by straight boundaries in a stability diagram displaying $\gamma e/\Gamma_c^0$ {\it v.s.} $e/\Gamma_i^0$ (fig.~\ref{Fig:DiagPhase}b).

However, the new collaborative mode of failure we have described may also propagate below the critical residual stress required for the propagation of isolated channel cracks.
This mode is experimentally observed in the case of moderate adhesion  and leads to the formation of patterns characterized by a robust width~\cite{Marthelot14}.
We have described this width in the case of parallel strips and eq.~\ref{eq:regime_permanent} derived from Griffith's energy balance provides a condition for the propagation of such cracks.
This new boundary corresponding to $E_f=E_{sc}$, $\nu_f=0.25$ and $h_f=h_{sc}$ is plotted in fig.~\ref{Fig:DiagPhase}b as a red curve.
In agreement with our experimental data, this boundary is significantly below the boundary for isolated channel cracks as $e$ becomes slightly larger than $\Gamma_i^0$.

Several failure modes compete in the sector of the diagram where $e>\Gamma_i^0$ and $e>\Gamma_c^0/2\gamma$.
Complex fracture patterns such as spiral and oscillatory paths resulting from ``follower cracks'' can indeed be observed~\cite{Marthelot14} in addition to the duos described in the present study or to the more classical classical channel cracks followed by delamination (Fig.~\ref{Fig:MTEOS}).
The selection between these different mechanisms certainly relies on the propagation dynamics of fracture and delamination and remains a challenging problem.
We hope that this open question will motivate further investigations.

\section{Conclusion}
\label{section:concl}
Recent studies have put in evidence a novel mode of failure of thin films, involving the collaboration between fracture and delamination~\cite{Marthelot14}.
This mode leads to various patterns displaying a characteristic width curiously robust~\cite{Sendova03, Meyer04, Lebental07, Wan09, Wu13}.
The present study focuses on a novel fracture pattern that is crucial for the understanding of fracture propagation in the presence of delamination: a pair of parallel cracks connected by a straight delamination front that propagates simultaneously and leads to the detachment of strips with a steady width. \\
We started by a quantitative description of our experimental observations conducted with a model system based on a standard sol-gel coating process.
We then interpreted and predicted the stable width observed in the experiments through two complementary approaches.
We first used the principle of local symmetry based on numerical computations of stress intensity factors.
In addition to this numerical method, we developed a physical model involving analytical estimations of the elastic energy.
Both approaches lead to a steady width, which is only determined by the elastic properties of the film (and the mismatch with the mechanical properties of the substrate). 
Strikingly, this width is independent from fracture and delamination energies.  \\
This robust strip width can be interpreted through the physical model as a balance between surface and line effects in the vicinity of the delamination front.  
Indeed, boundary conditions limit the relaxation of the residual stress in the strip while tilting the strip releases some additional elastic energy.
This model also provides an interesting prediction for the evolution of the strip width as a function of Poisson ratio of the film, and as a function of substrate compliance which are confirmed by numerical computation of stress intensity factors. \\
Although our study is limited to strip patterns, our approach could in principle be extended to more complex geometries such as spirals or alleys that are also observed in experiments.
However, computing stress intensity factors would be more difficult since the direction of the delamination front would have to be determined as well.
Conversely, the physical model, although approximate, still brings some insights~\cite{Marthelot14}.
This collaborative mode of failure is still far from being comprehensively explored. In particular, a better description of the dynamics is necessary to predict the selection of the failure mode. We nevertheless hope that our different approaches will open further investigations into more complex configurations. 
For instance, the phenomenon reported here could be captured by a numerical implementation of a damage model for a simplified system~\cite{Baldelli13, baldelli14} involving both fracture and delamination.

\section{Acknowledgements}
We thank J\'er\'emie Teisseire, Davy Dalmas and \'Etienne Barthel for help with the experiments and fruitful discussions.
We thank John W. Hutchinson, Corrado Maurini, Andr\'es A. Le\'on Baldelli and Jean-Jacques Marigo for very stimulating ideas and criticisms.
We thank ECOS C12E07, CNRS-CONICYT, and
Fondecyt Grant No. 1130922 for partially funding the project. 

\section{Bibliography}
\bibliography{references}
\bibliographystyle{unsrt}

\end{document}